# *New univariate characterization of fish community size structure improves precision beyond the Large Fish Indicator*

Christopher P. Lynam[1] and Axel G. Rossberg[1,2]

[1]Centre for Environment, Fisheries and Aquaculture Science, Pakefield Road, Lowestoft Suffolk NR33 0HT, UK

[2]School of Biological and Chemical Sciences, Queen Mary University of London, Mile End Rd, London E1 4NS, UK

20/07/17

**Abstract:** The size structure of fish-communities is an emergent high-level property of marine food webs responsive to changes in structure and function. To measure this food web property using data arising from routine fisheries surveys, a simple metric known as Typical Length has been proposed as more suitable than the Large Fish Indicator, which has been highly engineered to be responsive to fishing pressure. Typical Length avoids the inherent dependence of the Large Fish Indicator on a parameter that requires case-by-case adjustments. Using IBTS survey time series for five spatial subdivisions of the Greater North Sea, we show that the Typical Length can provide information equivalent to the Large Fish Indicator when fishing is likely the strongest driver, but differences can also arise. In this example, Typical Length exhibits smaller random fluctuations ("noise") than the Large Fish Indicator. Typical Length is also more adaptable than the Large Fish Indicator and can be easily applied to monitor pelagic fish in addition to demersal fish, and together with information on the potential growth of the fish community, a proxy of which can be derived from the Mean Maximum Length indicator, it is possible to partition change in community composition from change in size structure. This suggests that Typical Length is an improvement over the Large Fish Indicator as a food web indicator with the potential to offer further insight when considered in conjunction with indicators of community composition.

**Introduction**

The Large Fish Indicator (LFI) is a well established management indicator for the status of marine demersal fish communities is (Greenstreet et al. 2011; OSPAR 2017a). It is defined as the proportion by biomass of fish longer than a threshold length $L_{\mathrm{large}}$, that are caught in a given scientific survey, where $L_{\mathrm{large}}$ is a parameter of the indicator. Depending on circumstances, $L_{\mathrm{large}}$ needs to be adjusted (Greenstreet et al. 2011) to maintain the high sensitivity characteristic of the LFI. Initially, LFI time series were computed only for the North Sea with $L_{\mathrm{large}} = 40\mathrm{cm}$ and this parameter dependence was unproblematic. Later, however, the indicator's success led to its application in many other contexts, e.g. the deep sea (Mindel et al. 2016), pelagic communities (Oesterwind et al. 2013), or marine shelf regions with exploitation histories different from the North Sea (Shephard, Reid, and Greenstreet 2011). In such cases, adjustments of $L_{\mathrm{large}}$ have been proposed or appear plausible for obtaining informative indicator time series. Indeed, the threshold length for the North Sea IBTS has recently been adjusted to 50cm given an update to the time series information and different values has been chosen for other surveys in the same area with the acknowledgement that the current method was unable to find the optimum signal-to-noise ratio in each circumstance (OSPAR 2017a). Such adjustments, raise two issues. Firstly, LFI values computed for different $L_{\mathrm{large}}$ thresholds cannot easily be compared. This limits the usefulness of the LFI for any situation where contrast in the size structure of fish communities are large. *De novo* application of proposed procedures to find the "ideal" value for $L_{\text{large}}$ for a given survey time series (Shephard, Reid, and Greenstreet 2011), which aims to increase indicator signal and minimize noise, carries the risk of introducing biases or "overfitting" that generates indicator signals stronger than warranted by the biology of the study system. Engelhard et al (2015) and Stamoulis et al. (2016) show clear spatial structure in the LFI in the North Sea, suggesting that the value of $L_{\text{large}}$ should not be constant within the North Sea and community structure, when addressed spatially, will likely result in differing values between communities even when measured by the same survey.

As a potential replacement for the LFI that overcomes these difficulties while retaining the high management utility of the LFI, we study here the indicator called Typical Length (OSPAR 2017b). The indicator has originally been proposed in a working group report by the International Council for the Exploration of the Sea (ICES 2014) and considered alongside the LFI by OSPAR (2017), but so far little is known about the statistical properties of TyL in comparison to LFI. The present note aims to contribute to a better understanding of this aspect.

**Materials and Methods**

*Definition of Typical Lengths and its motivation*

Typical Length (TyL) is defined as the geometric mean length of fish, weighted by biomass. In other words, if $M_i$ is the body mass of the $i$-th fish called in a given survey area and year, and $L_i$ its length, and if a total of $N$ individuals have been caught, then

$$\text{TyL} = \exp\left[\frac{\sum_{i=1}^{N} M_i \ln L_i}{\sum_{i=1}^{N} M_i}\right].$$

The units in which TyL is measured are identical to those in which the lengths $L_i$ are measured.

As explained by ICES (2014), an advantage over many other length-based indicator that TyL has in common with the LFI (with appropriate $L_{\text{large}}$) is that the formula for the indicator value gives approximately equal statistical weight to all length classes of fish on a logarithmic axis. Since the biomass of marine communities tends to be approximately evenly distributed over the logarithmic length axis (Sheldon, Prakash, and Sutcliffe, Jr. 1972), this balances the sensitivity of the indicator over the entire fish community. Other length-based indicators give either disproportionally strong weight to the smallest size classes (e.g. the arithmetic mean length of individuals) or the largest size classes (e.g. 95$^{\text{th}}$ percentile of the length distribution of individuals weighted by biomass) (ICES 2014). When some size classes have particularly strong weight in the computation of an indicator, relatively small fluctuation or measurement errors in these size classes could be detrimental to the indicator's signal to noise ratio. This is the problem that, based on the structure of their definitions, both LFI and TyL aim to avoid.

*Data*

The dataset of fish size and abundance using in this study derives from International Bottom Trawl Surveys (IBTS) in the North Sea. It has been carefully checked for errors and corrected where necessary by Marine Scotland Science, who kindly provided this data set to us.

As shown in Fig. 1, this data has consistent high coverage of the wider North Sea and its five subdivisions proposed by ICES (2015): Kattegat and Skagerrak (KS), and North Easter (NE), North

Western (NW), South Eastern (SE), and South Western (SW) North Sea. On average, each IBTS statistical rectangles (STSQ) is sampled in two hauls per year.

*Statistical Analysis*

For each subdivision, we compute times series of the three indicators LFI, TyL, and MML.

We used two different methods to estimate errors (the "noise") in indicator values. The first is was to compute a LOESS smooth the raw indicator time series using the function `loess` of R (version. 2.15.2012-09-19) with default parameters. The output of this function also provides standard errors for the fitted curves, one form of error estimate that we used.

As a second method to estimate errors in the indicator values we re-sampled the indicator time series 199 times using the following scheme: for each haul in the dataset a replacement is chosen either from the STSQ in which it was made (if there were replicates) or from one of the eight adjacent STSQs. Candidate hauls are selected from the cluster of surrounding hauls (from nine STSQs) with equal probability each time i.e. with replacement. This strategy tends to preserve the spatial structure of the sampling, eliminate edge effects and minimise the loss of rectangles from the surveyed area. For each bootstrap (i.e. when all hauls have been considered for replacement for each year), the three indicators are computed for each of the five subdivision. The 2.5% and 97.5% quantiles of the resulting 199 realisation of each data point are then used to estimate 95% confidence intervals.

**Results**

In Figure 2 we compare the time series of the three indicators for the five subdivisions, including error estimates obtained using the two methods described above. Strikingly, the LOESS smoothes for LFI and TyL behave very similar (up to some linear transformation) for the majority of comparisons with the exception of NW, where TyL picks up a strong recruitment event but LFI screens this out, and in the last few years of the SW comparison, where TyL suggests a recent improvement but LFI continues to decline. The time series for MML also captures the decline of fish-community size structure during the mid 1980s displayed by LFI and TyL, but fails to indicate the recent recoveries of size structure in KS, NW, and NE that LFI and TyL demonstrate suggesting that although size-structure is recovering the community composition is still dominated by species that do not attain a large size. In comparison with the other two indicators the LOESS smooth of MML is more "wavy", possibly a reflection of strong recruitment events of species that, although they could potentially grow large, in practice suffer high mortality at small size .

Figure 3 demonstrates the differing $L_{\text{large}}$ thresholds that could be applied for communities in each area and the effect that this would have on the resulting LFI time-series. Notably, very high signal to noise can be obtained by virtually eliminating the signal such that LFI scores are near zero.

When comparing the error estimates for LFI and TyL, it is visually clear that those for TyL tend to be slightly smaller than those for LFI, for both methods considered: LOESS smooth and re-sampling.

**Conclusions and future development**

Our simple analysis shows that TyL not only overcomes the problems resulting from the parameter dependence of LFI described above, TyL is also slightly less "noisy" than LFI in this comparison. Otherwise, the information provided by the two indicators (the "signal") is similar in this system, where fishing has been the overwhelming pressure on demersal fish communities. However, further study is required to determine whether the two indicators (LFI and TyL) will behave differently as the system becomes driven by natural mortality and begins to recover, as is expected in the North Sea under the current low fishing effort regime. Further comparisons will need to be considered.

At this stage, the LFI has merit as an indicator sensitive to demersal fishing pressure, but we can recommend that the TyL indicator be considered as an additional indicator for food web assessments, particularly since the data requirements of the two are equal. To improve the TyL as a food web indicator we will incorporate survey efficiency corrections (Walker et al 2017) so that the biomasses of species in the indicator more closely resemble their true biomass in the system. Additionally, the responses of TyL for pelagic fish communities and for combined pelagic and demersal communities should be investigated through analytical and modelling studies. Once this further work is complete, the TyL indicator may prove to much more responsive to food web functioning than the LFI, but less responsive to demersal fishing pressure. Alternatively, TyL may prove to be more responsive to fishing and serve as a full replacement for the LFI.

By contrast, the MML time series does not show some characteristic features of the LFI time series, in particular the recent recovery in size-structure in parts of the North Sea. However, the complementary information provided by MML is valuable, since the indicator responds clearly to change in species composition, and deserves further consideration by management. We believe that the good ability of both LFI and TyL to indicate varying degrees of success of recent changes in fisheries policy speak for their use in policy contexts.

**Acknowledgements:** C.P.L. acknowledges funding by the UK Department of Environment, Food and Rural Affairs (MF1228; "Fizzyfish: from Physics to Fisheries") and A.G.R. by the Natural Environment Research Council and the UK Department for Food, Environment and Rural Affairs (NE/L00299X/1; MERP: Marine Ecosystems Research Program).

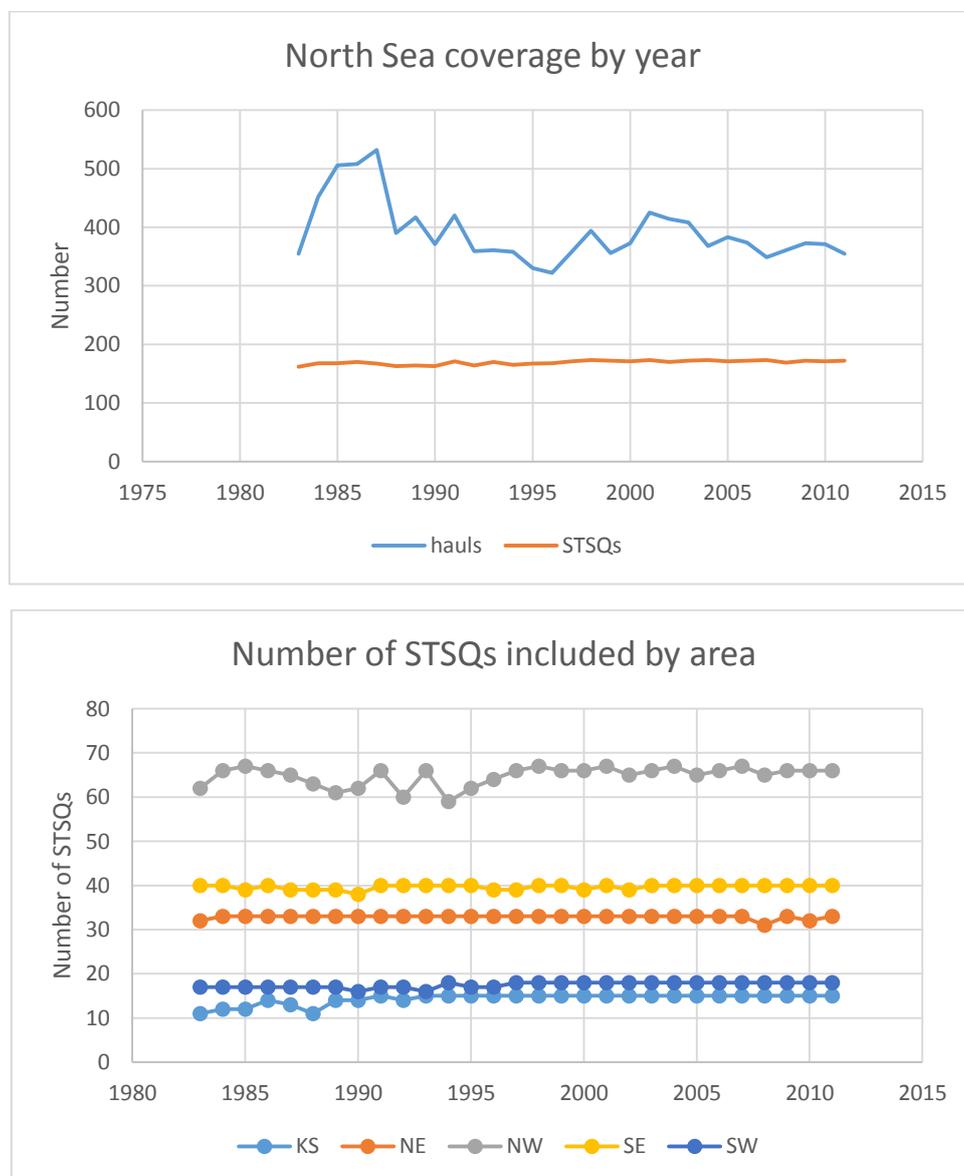

Figure 1. Coverage of the North Sea through time and space by our data set. STSQ = The number of IBTS statistical rectangles (or "squares") covered.

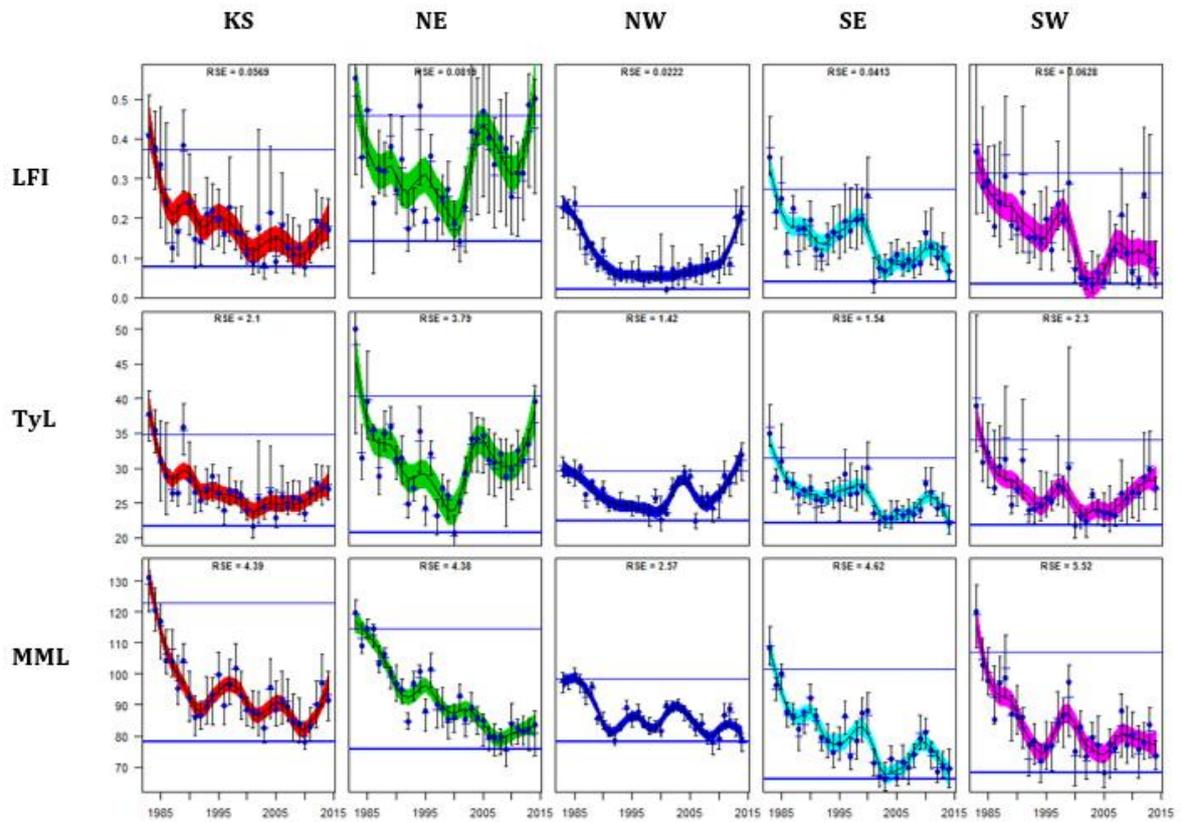

Figure 2. Time series for the three indicators LFI, TyL and MML in five North Sea sub-regions, with error bars indicating 95% confidence intervals from bootstrapping, the solid line the LOESS smooth, and the coloured error the $\pm 1$ SE band around this smooth.

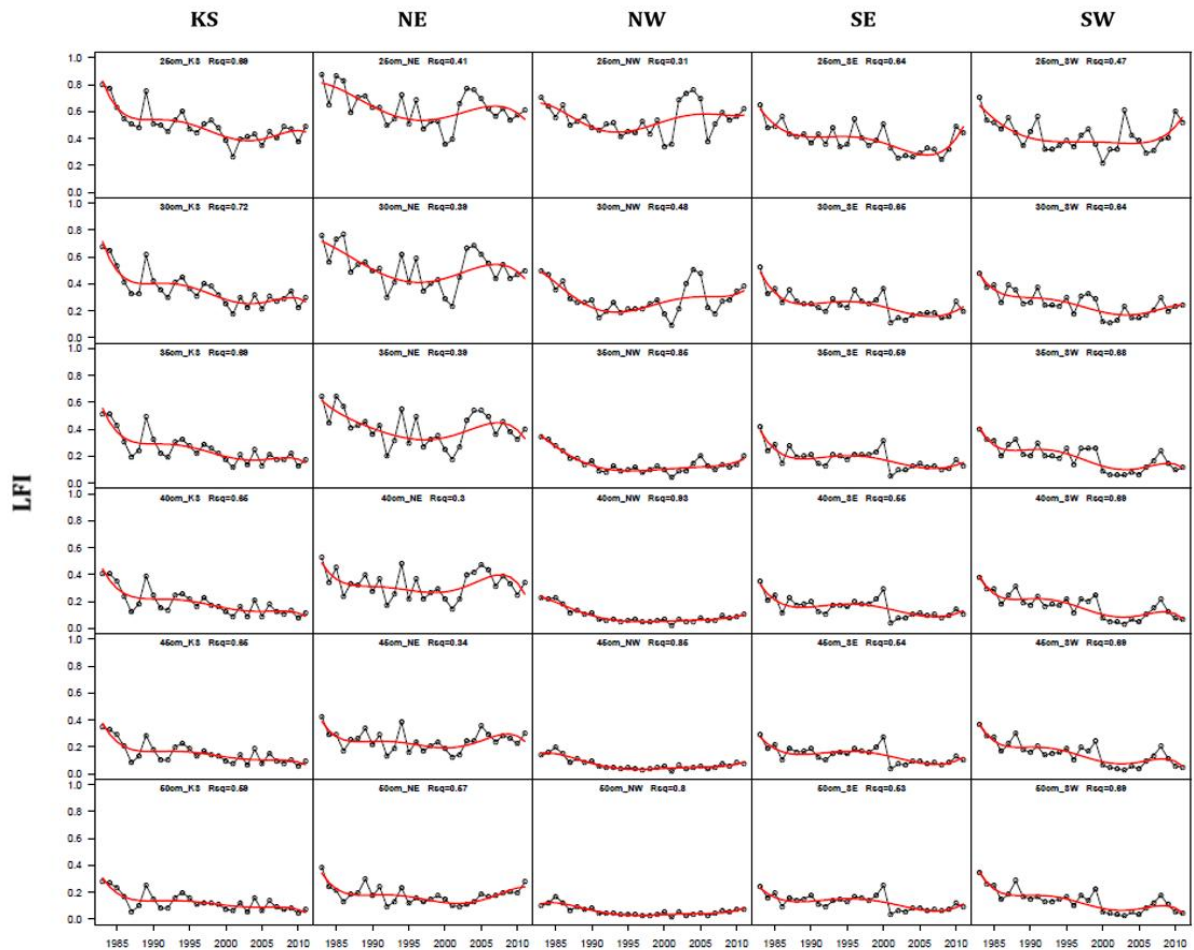

**Figure 3. LFI for demersal fish by sub-region.** The large fish thresholds tested are from 25 cm (top) to 50 cm (bottom) in steps of 5 cm. The variability explained by the polynomial fit is given in the subtitles by the $R^2$ values.